\pdfoutput=1

\documentclass[journal]{IEEEtran}
\ifCLASSINFOpdf
\else
\fi
\usepackage{lineno,hyperref}
\usepackage{color}
\usepackage{subfigure}

\usepackage{cite}
\usepackage{amsmath,amssymb,amsfonts}
\usepackage{algorithmic}
\usepackage{graphicx}
\usepackage{textcomp}

\begin{document}
%

\title{Riemannian Nearest-Regularized Subspace Classification for Polarimetric SAR images}
%
%
%

\author{Junfei Shi,
        Haiyan Jin\vspace{-2em}
\thanks{Junfei Shi was with  the Department
of Computer Science and Technology, Shaanxi Key Laboratory for Network Computing and Security Technology, Xi'an University of Technology, Xi'an, China.
corresponding author: shijunfei@xaut.edu.cn.}
\thanks{Haiyan Jin was the professor with the Department
of Computer Science and Technology, Shaanxi Key Laboratory for Network Computing and Security Technology, Xi'an University of Technology, Xi'an, China.}
\thanks{This work was supported in part by the National Natural Science Foundation of China under Grant 62006186, the Science and Technology Program of Beilin District in Xi'an under Grant GX2105,
in part by the Open fund of National Key Laboratory of Geographic Information Engineering under Grant SKLGIE2019-M-3-2.}}

\maketitle

\begin{abstract}
As a representation learning method, nearest regularized subspace (NRS) algorithm is an effective tool to obtain both
accuracy and speed for PolSAR image classification. However, existing NRS methods use the polarimetric feature vector but the PolSAR original covariance matrix (known as  Hermitian positive definite (HPD) matrix) as the input. Without considering the matrix structure, existing NRS-based methods cannot learn correlation among channels. How to utilize the original covariance matrix to NRS method is a key problem. To address this limit, a Riemannian NRS method is proposed, which consider the HPD matrices endow in the Riemannian space. Firstly, to utilize the PolSAR original data, a Riemannian NRS method(RNRS)  is proposed by constructing HPD dictionary and HPD distance metric. Secondly, a new Tikhonov regularization term is designed to reduce the differences within the same class. Finally, the optimal method is developed and the first-order derivation is inferred. During the experimental test, only T matrix is used in the proposed method, while multiple of features  are utilized for compared methods. Experimental results demonstrate the proposed method can outperform the state-of-art algorithms even using less features.
\end{abstract}

\begin{IEEEkeywords}
Riemannian Nearest-Regularized Subspace(RNRS), PolSAR image classification, Riemannian metric.
\end{IEEEkeywords}

%
\IEEEpeerreviewmaketitle

\section{Introduction}
%
%
%
%
\IEEEPARstart{T}{he} fully polarimetric synthetic aperture radar (PolSAR)\cite{2019A} image classification is an important task for radar image processing. Compared with SAR system, PolSAR system can obtain four channel PolSAR data  since it emits and receives electromagnetic waves with two orthogonal polarization modes. Thus, more target scattering characteristics can be provided  for image classification. In recent years, PolSAR system have been widely used in image classification\cite{2019A}, target recognition\cite{2020Object}, land cover mapping and so forth. Existing PolSAR image classification methods including unsupervised and supervised methods have been widely put foreword.

Recently, representation-based classification methods have been demonstrated to be an effective tool in the field of radar image processing\cite{8994163}. Representation classification methods can be mainly concluded into three categories: sparse representation classification(SRC)\cite{8486711}, collaborative representation classifier(CRC)\cite{8716570} and nearest regularization subspace(NRS)\cite{6472065,Zhang2017Nearest,2019Robust} classification.
SRC has been widely used in terrain classification of hyperspectral and SAR images. In Ref. \cite{6918378}, the  SRC is extended to PolSAR image by combining polarimetric features and sparse representation. In \cite{7947120}, Remianian sparse coding is proposed for PolSAR images. All these methods have the regularization term with l0- or l1- norm, which is non-linear NP-hard problem. And the iterative resolving method, such as Simultaneous OMP (SOMP) \cite{1416405}, is time-costed. Collaborative representation classifier(CRC)\cite{7729793} considers the main contribution in SRC is not sparse but collaborative. So, the l2-norm regularization is used in the CRC method instead of l1-norm. Thus, the CRC model is convex optimization problem which reduces the computing time greatly.

NRS methods\cite{6472065,Zhang2017Nearest,2019Robust}, as the extended version of collaborative representation classifier(CRC), is another popular representation-based method for PolSAR image classification. Compared with CRC, NRS method minimizes the intra-class distance by adding the Tikhonov matrix regularization. In \cite{6472065}, the NRS method is proposed for hyperspectral image classification. In \cite{Zhang2017Nearest}, the NRS method is applied to PolSAR image classification by using polarimetric feature vector and spatial information. In \cite{2019Robust}, a robust weighting nearest regularized subspace classifier is designed for PolSAR imagery.  Compared against SRC, the regularization term in the NRS is l2-norm, which has the closed-form solution and can reduce the computing time greatly. In addition, compared with CRC, the Tikhonov regularization can suppress the speckle noises and improve the classification accuracy. Hence, the NRS is the optimal model to balance the accuracy and effectiveness.

However, PolSAR data is a complex matrix, known as the HPD matrix. The NRS method cannot learn PolSAR data well, since it can only process the feature vector but not the HPD matrix.
Generally, the NRS method converts the PolSAR covariance matrix into a column vector as the input. However, matrix vectorization cannot preserve the data structure and cross-polarization information. Moreover, the column vector destroys the matrix distribution and instinct correlation among channels. Besides, some other methods use the extracted target feature vector as the input, while the classification performance strongly depends on what kinds of features are extracted. How to choose the right features is a challenging practical problem. So, why we cannot use the original covariance matrix directly as the input, and how to utilize the original matrix to the NRS method is a natural question. In order to address these problems, we proposed the Riemannian NRS method, in which the original C matrix is adopted to construct the data dictionary for each class. The C matrix is an HPD matrix, which endows to a Riemannian manifold space\cite{Anoop2017Riemannian}. Based on this, a new metric distance and novel Riemannian NRS model should be built.

In this letter, we propose the Riemannian NRS method for PolSAR classification. The main contributions of our work can be summarized into three aspects. 1) To learn the PolSAR data effectively, the original C matrix is used as the input and formulate the dictionary, and the Riemannian Nearest-Regularized Subspace(RNRS) classification model is proposed by utilizing the geometric distance of the HPD matrix. 2) To further reduce the speckle noises, a new Tikhonov matrix is defined to minimize the intra-class differences within the same class. 3) A fast optimal method is  inferred to resolve the RNRS model with a first-order method. In a word, in this letter, the original C matrix is utilized, and the Riemannian metric is extended to the NRS for the first time. And the matrix structure and channel information can be fully exploited.

This letter is organized as follows. The PolSAR data and Riemannian metric are introduced in Section II.  The proposed method is given in Section III. Section  IV  is the experimental study. The conclusion is given in Section  V.


\section{PolSAR data and Riemannian metric}

The PolSAR data contain more scattering information than traditional SAR data, since its scattering echoes are from four channels according to different emitting and receiving modes. The polarimetric data can be expressed as S matrix:

\begin{equation}\label{1}
S = \left[ {{S_{hh}},{S_{hv}},{S_{vh}},{S_{vv}}} \right]
\end{equation}
where $h$ and $v$ are the horizontal and vertical emitting and receiving modes respectively.

Under the condition of satisfying the theory of reciprocity, $S_{hv}=S_{vh}$. The S matrix can be vectored as $\emph{\textbf{k}} = {\left[ {{S_{hh}},\sqrt 2 {S_{hv}},{S_{vv}}} \right]^T}$ under the Pauli base. Through multi-look processing, each pixel is commonly represented as the covariance matrix C or coherency matrix T, which are the most frequency used data formats during the PolSAR image processing.

It is widely known that the polarimetric covariance matrices are Hermitian (semi-)positive definite (HPD), which form a Riemannian manifold\cite{7947120} but not Euclidean space. The similarity of two points in Riemannian manifold is measured by the Riemannian metric, which can be extended to PolSAR data space. Here we use the well-known metric distance: affine invariant Riemannian metric (AIRM)\cite{IJCV}. For two points in PolSAR image, the AIRM distance of two HPD matrices X and Y can be defined as:
\begin{equation}\label{5}
{d_R}\left( {X,Y} \right) = {\left\| {\log ({X^{ - \frac{1}{2}}}Y{X^{ - \frac{1}{2}}})} \right\|_F}
\end{equation}
%
%


 AIRM is affine invariant in a curved geometry space with some  properties, e.g., invariant to affine transformation\cite{6856946}. This distance is defined for SPD matrices, and their geometry naturally extends to the HPD case \cite{Anoop2017Riemannian}. The AIRM is the intrinsic Riemannian metric that corresponds to a geodesic distance on the manifold of HPD matrices.
\section{Proposed Method}

\subsection{Riemannian Nearest-Regularized Subspace(RNRS) Model}

Nearest-Regularized Subspace for vector data has achieved great success and been widely used in image classification\cite{Zhang2017Nearest,2019Robust}, in which each pixel is expressed as a feature vector. In the NRS, dictionary is composed of a set of feature vectors by selecting samples randomly for each class. NRS assumes that a test pixel can be estimated by the linear combination of the given dictionary for each class in European space. The pixel is assigned as the class label with the minimum residue between the test sample and the estimation one. This approach achieves superior performance in image classification. In \cite{Zhang2017Nearest}, the NRS is extended to PolSAR classification, in which the covariance matrix is pulled into a column vector. However, it  not only destroys the matrix structure but also the polarimetric information among different PolSAR channels.

The PolSAR data is an HPD matrix instead of a vector, and its distribution is under the Riemannian space instead of the European space. According to the data characteristics, we try to propose the Riemannian Nearest-Regularized Subspace(RNRS) method for PolSAR images. The dictionary should be constructed from a set of HPD-valued data, and the dictionary element is the polarimetric covariance matrix instead of the column vector. Therefore, each HPD-valued pixel can be linear combination of other dictionary elements for each class in Riemannian space.

Assume the HPD-valued dictionary set $D = \{ {D^1},{D^2}, \cdots ,{D^C}\}$, and $C$ is the total class number. ${D^k}$ is $k$-th class sub-dictionary composed of $N_k$ atoms $\left\{ {D_1^k,D_2^k, \cdots ,D_{{N_k}}^k} \right\}$, where$D_i^k$ is an HPD covariance matrix from training samples of class $k$. For a test matrix X, the main goal of RNRS method is to find its most similar estimated value $\bar X$, and assigned the corresponding class label to X. The estimated value can be achieved by linear combination of dictionary atoms for each class, defined as

\begin{equation}\label{13}
{\overline X _k} = {D^k}{\alpha _k} = \sum\limits_{i = 1}^{{N_k}} {D_i^k\alpha _i^k} ,k = 1 \sim C
\end{equation}

$D_i^k$ is the $i$th dictionary atom for class $k$, $\alpha_i^k$ is the coefficient of the $i$th atom. So, ${\alpha _k}{\rm{ = \{ }}\alpha _1^k,\alpha _2^k, \cdots ,\alpha _{{N_k}}^k\}$ is the column vector of coefficients, which can be calculated by

\begin{equation}\label{14}
\alpha _k^* = \mathop {\min }\limits_{{\alpha _k}} {\rm{ }}\phi {\rm{(}}\alpha {\rm{) = }}\left\| {d(X,{{\bar X}_k})} \right\|_2^2 + \lambda \left\| {\Gamma _X^k{\alpha _k}} \right\|_2^2
\end{equation}
where $\lambda$ is the weight factor to balance the residual and coefficient items. $d(X,{\bar X_k})$ measure the Riemannian geodesic distance between X and ${\bar X_k}$. Here we use the AIRM Riemannian geodesic distance. Equ. (\ref{14}) can be rewritten as:

\begin{equation}\label{15}
\alpha _k^* = \mathop {\min }\limits_{{\alpha _k}} {\rm{ }}\phi {\rm{(}}\alpha {\rm{)}} = \left\| {\log (\sum\limits_{i = 1}^{{N_k}} {\alpha _i^k{X^{{\rm{ - }}\frac{1}{2}}}D_i^k{X^{{\rm{ - }}\frac{1}{2}}}} )} \right\|_F^2 + \lambda \left\| {\Gamma _X^k{\alpha _k}} \right\|_2^2
\end{equation}

Moreover, $\Gamma _X^k$ measure the Riemannian geodesic distance between test sample X and dictionary atoms in the $k$th class, defined as:

\begin{equation}\label{16}
\Gamma _X^k = \left[ {\begin{array}{*{20}{c}}
{\Gamma _{X,1}^k}& \cdots &0\\
 \vdots & \ddots & \vdots \\
0& \cdots &{\Gamma _{X,{N_k}}^k}
\end{array}} \right]
\end{equation}

\begin{equation}\label{17}
\Gamma _{X,i}^k = \left\| {d(X,D_i^k)} \right\|_2^2 = \left\| {\log ({X^{ - \frac{1}{2}}}D_i^k{X^{ - \frac{1}{2}}})} \right\|_F^2
\end{equation}

$\Gamma _X^k$ compute the distance between the pixel X and each atom in class $k$, and $\Gamma _{X,i}^k$ is defined as the Riemannian geodesic distance. Thus, the second regularization term in Equ.(\ref{15}) can reduce the effect of speckle noises by increasing the weights of dictionary atoms that are close to the pixel X.

Finally, the class label of the test sample can be determined by the residual minimum values among all the classes:

\begin{equation}\label{18}
class(X) = \mathop {\min }\limits_k {\rm{ }}{d^2}(X,{\bar X_k}),{\rm{  }}k = 1 \sim C.
\end{equation}

\subsection{Optimization Method}

\textbf{\emph{Optimization equation (\ref{15})}}: To optimize (\ref{15}), we divide it into two parts. The first part is the residual error term, and the second one is the Tikhonov regularization term. So, the objective function can be written as:
\begin{equation}\label{19}
\min f\left( \alpha  \right) = {f_1}\left( \alpha  \right) + {f_2}\left( \alpha  \right)
\end{equation}
where
\begin{equation}\label{20}
 {f_1}\left( \alpha  \right) = \left\| {\log (\sum\limits_{i = 1}^{{N_k}} {\alpha _i^k{X^{{\rm{ - }}\frac{1}{2}}}D_i^k{X^{{\rm{ - }}\frac{1}{2}}}} )} \right\|_F^2
\end{equation}

\begin{equation}\label{21}
 {f_2}\left( \alpha  \right) = \lambda \left\| {\Gamma _X^k{\alpha _k}} \right\|_2^2
\end{equation}

A first-order method is used to optimize them. First, we try to find the first-order derivation of the error term. According to the F-norm defination,${\left\| Z \right\|_F} = Tr\left( {{Z^T}Z} \right)$. The first-order derivation process is given as follows.

Assume $M\left( \alpha  \right){\rm{ = }}\sum\limits_{j = 1}^n {{\alpha _j}{D_j}}$, $S = {X^{ - \frac{1}{2}}}$. We have
\begin{equation}\label{22}
{f_1}\left( \alpha  \right) = Tr\left( {\log {{\left( {SM\left( \alpha  \right)S} \right)}^T}\log \left( {SM\left( \alpha  \right)S} \right)} \right)
\end{equation}

The chain rule of calculus then immediately yields the derivation:
\begin{equation}\label{23}
{{f'}_1}\left( \alpha  \right) = 2Tr\left( {\log \left( {SM\left( \alpha  \right)S} \right){{\left( {SM\left( \alpha  \right)S} \right)}^{ - 1}}SM'\left( \alpha  \right)S} \right)
\end{equation}
which simplifies as desired.

Given $M\left( {{\alpha _p}} \right) = {\alpha _p}{D_p} + \sum\limits_{i \ne p} {{\alpha _i}{D_i}}$  and according to the equation (\ref{23}), for the objective function ${f_1}\left( \alpha  \right)$, we can obtain
\begin{equation}\label{24}
\frac{{\partial {f_1}\left( \alpha  \right)}}{{\partial {\alpha _p}}} = Tr\left( {\log \left( {SM\left( {{\alpha _p}} \right)S} \right){{\left( {SM\left( {{\alpha _p}} \right)S} \right)}^{ - 1}}S{D_p}S} \right)
\end{equation}

Next, we give the first-order derivation of the Tikhonov regularization term. For objective function ${f_2}\left( \alpha  \right)$, we can obtain directly

\begin{equation}\label{25}
{f'_2}(\alpha ) = 2\lambda {\alpha _k}\Gamma _X^T{\Gamma _X}
\end{equation}
where ${\Gamma _X}$  can be calculated by equations (\ref{16}) and (\ref{17}).

Finally, the first-order derivation of objective function in equation (\ref{15}) can be obtained as:
\begin{equation}\label{26}
\begin{aligned}
\frac{{\partial \phi \left( \alpha  \right)}}{{\partial {\alpha _p}}} &= Tr\left( {\log \left( {SM\left( {{\alpha _p}} \right)S} \right){{\left( {SM\left( {{\alpha _p}} \right)S} \right)}^{ - 1}}S{D_p}S} \right) + \lambda \alpha \Gamma _X^T{\Gamma _X}\\
 &= Tr(\log (\sum\limits_{i = 1}^{{N_k}} {\alpha _i^k{X^{ - \frac{1}{2}}}{D_i}^k{X^{ - \frac{1}{2}}}} ){(\sum\limits_{i = 1}^{{N_k}} {\alpha _i^k{X^{ - \frac{1}{2}}}{D_i}^k{X^{ - \frac{1}{2}}}} )^{ - 1}}\\
&{X^{ - \frac{1}{2}}}D_p^k{X^{ - \frac{1}{2}}}) + \lambda {\alpha ^k}\Gamma _X^T{\Gamma _X}
\end{aligned}
\end{equation}

Although the equation (\ref{15}) is a non-negative convex function, the analytical solution cannot be resolved since the matrix derivation is complicated. So, an approximated solution of equation (\ref{15}) can be calculated by the spectral projected gradient (SPG) method\cite{2001SPG} using the first-order derivation by equation (\ref{26}).


\section{Experimental Study}

\subsection{Experimental Settings}

In this section, two sets of PolSAR images with different bands and radars are used to evaluate the performance of the proposed RNRS classification method. The first data set is an L band AIRSAR fully Polarimetric SAR image, which is acquired in Flevoland area, the Netherlands. Another one is an 4-look C band San Francisco
area fully polarimetric data from RADARSAT-2 with the resolution of 5m. For the two PolSAR images, the corresponding label maps are given as the reference for the quantitative evaluation. And some evaluation indicators, such as overall accuracy,  average accuracy, Kappa coefficient and confusion matrix, are calculated to testify the performance of the proposed method.

Moreover, three state-of-art methods are used for comparison. They are the NRS,  NRS\_MRF and weighted NRS methods respectively.
However, the three compared methods can hardly obtain good performance with only T matrix by the Euclidean distance. Here, we add 53 features including the T vector \cite{2020A} (shorted by ``3F+T" features) as the inputs of the three compared methods. While only the T matrix(shorted by ``only T" feature) is used as the input of the proposed RNRS method. Although it is unfair to our method,  the experimental results prove our method outperform others even with less features.

In addition, we randomly select 200 samples from labeled pixels for each class to formulate the dictionary for all the compared and proposed methods. The weight $\lambda$ is set to 0.1 by experiments. A computer with Intel Core i7 CPU and 64G RAM is used, and all the experiments are conducted on Window10 system with Matlab 2016a.

\subsection{Experimental Results of Flevoland Data Set}

\begin{figure*}
\centering
\setlength{\fboxrule}{0.2pt}
  \setlength{\fboxsep}{0.01mm}
\subfigure[]{\fbox{\includegraphics[height=0.1\textheight]{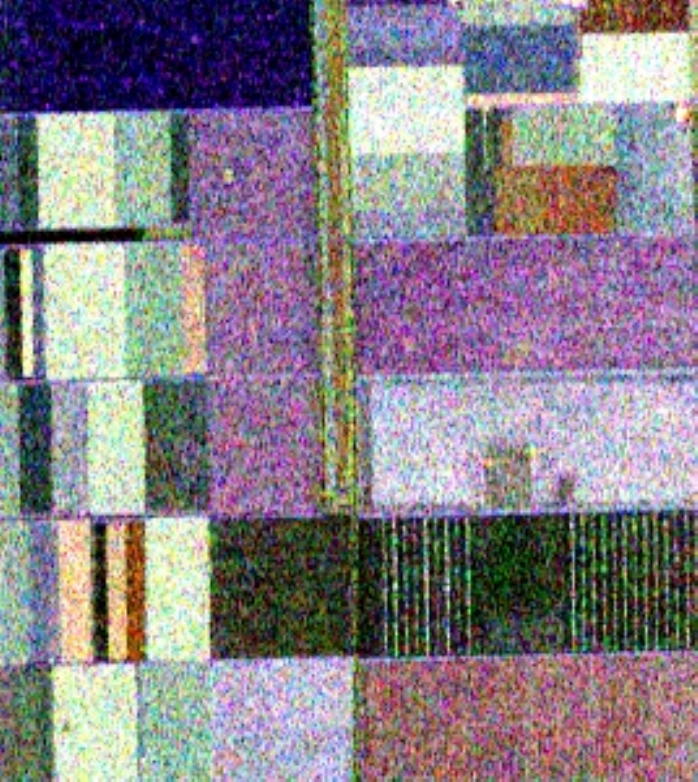}}}
\subfigure[]{\fbox{\includegraphics[height=0.1\textheight]{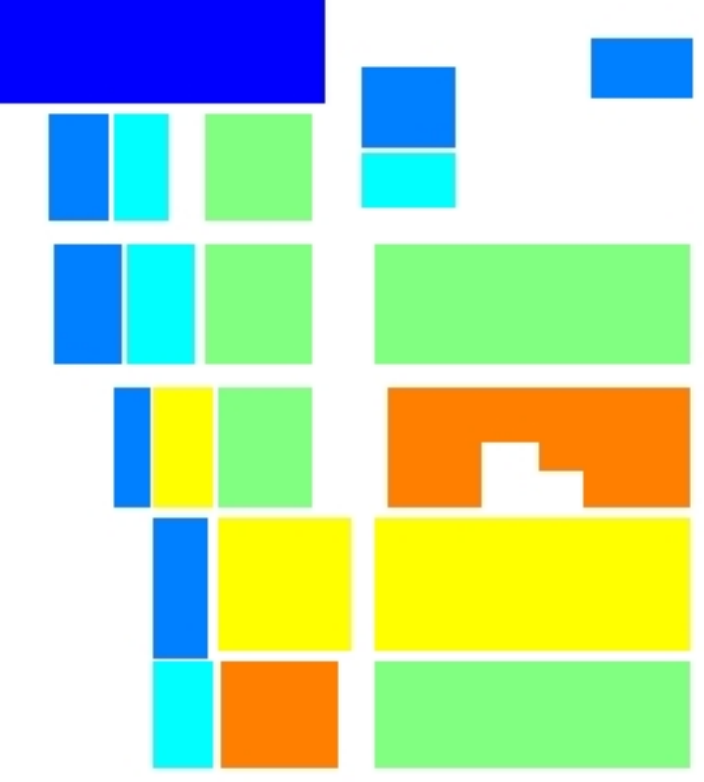}}}
\subfigure[]{\fbox{\includegraphics[height=0.1\textheight]{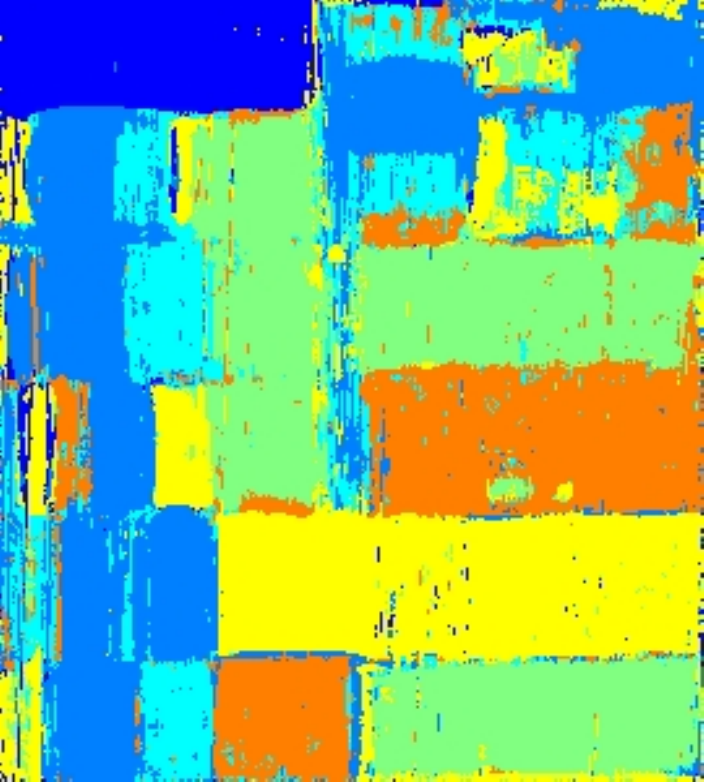}}}
\subfigure[]{\fbox{\includegraphics[height=0.1\textheight]{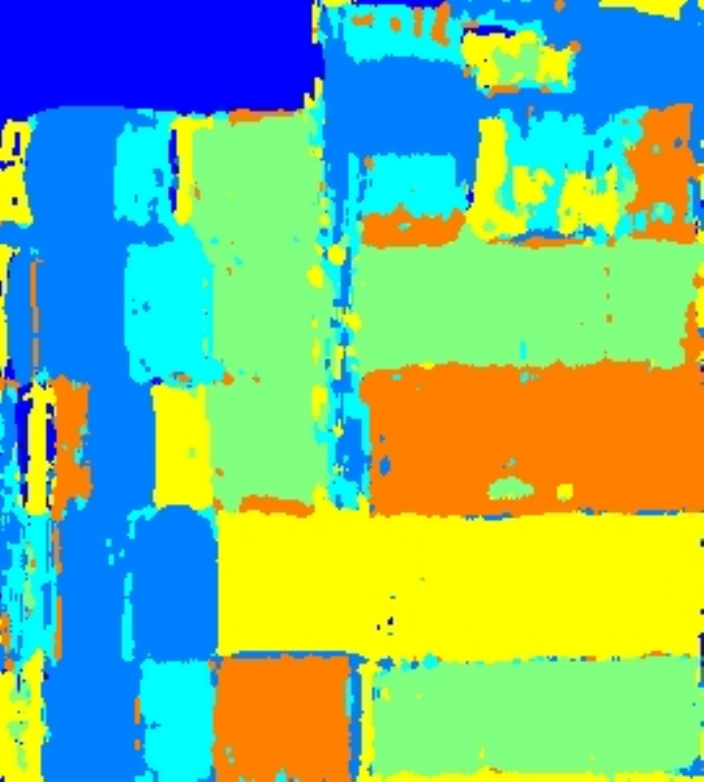}}}
\subfigure[]{\fbox{\includegraphics[height=0.1\textheight]{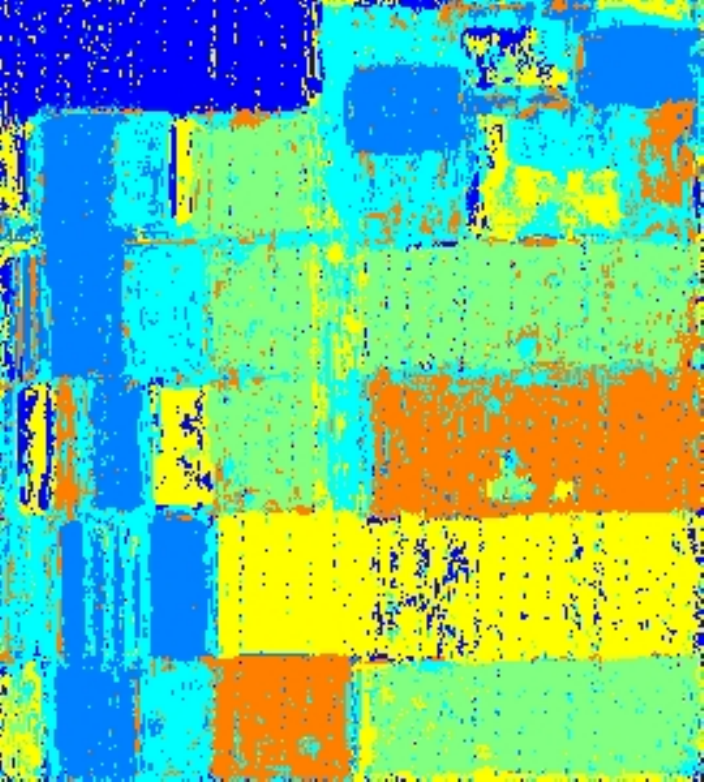}}}
\subfigure[]{\fbox{\includegraphics[height=0.1\textheight]{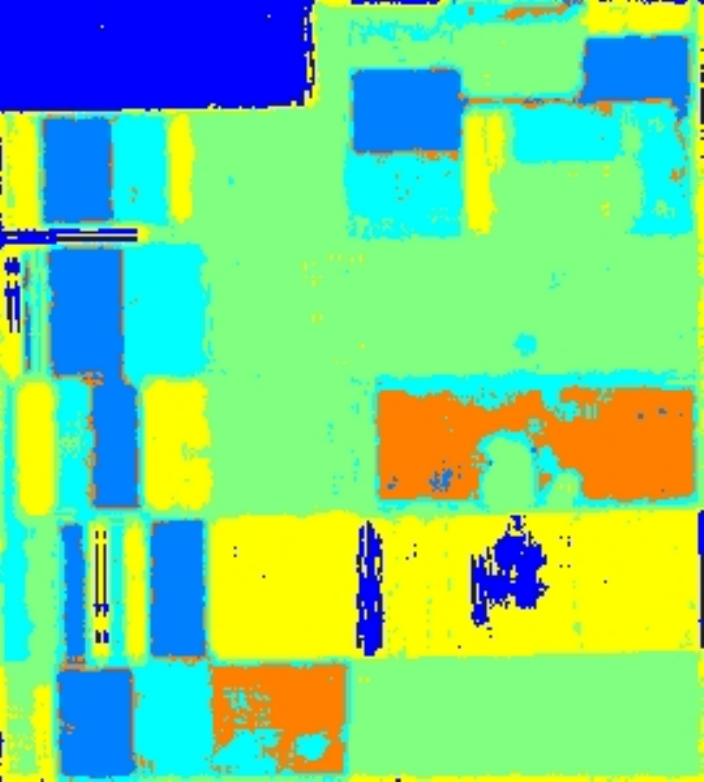}}}
\subfigure[]{\fbox{\includegraphics[height=0.1\textheight]{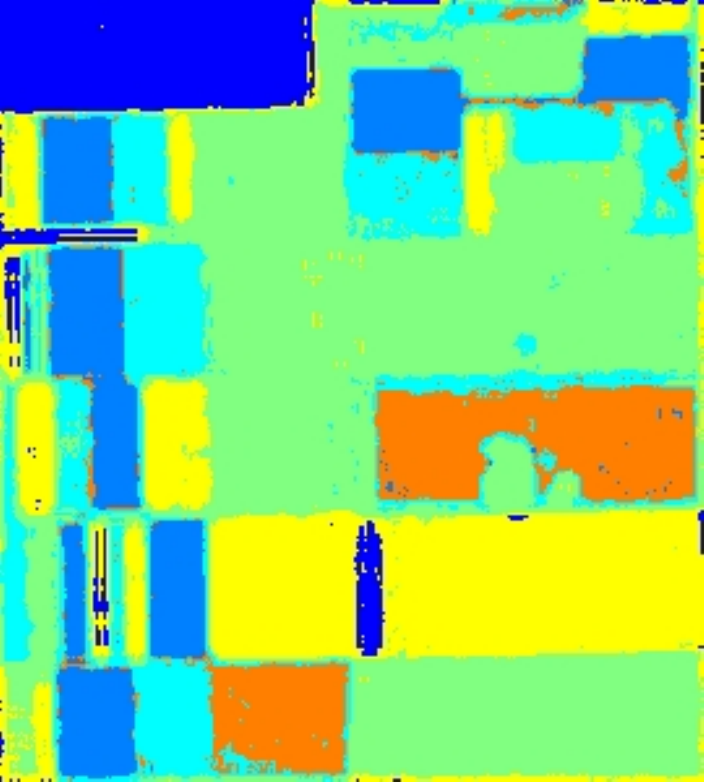}}}
\subfigure[]{\includegraphics[height=0.1\textheight]{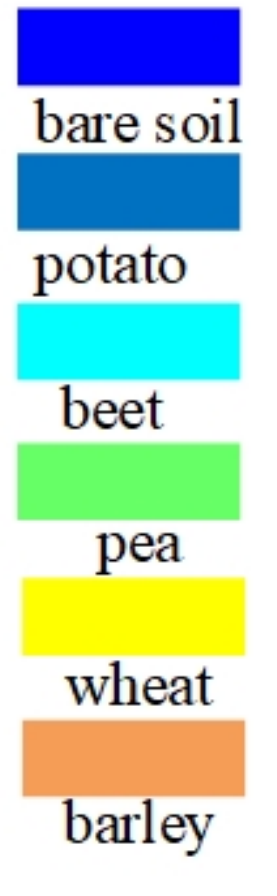}}

\caption{Classification results of FLEVOLAND data set. (a) PauliRGB image of FLEVOLAND area; (b) Ground truth of PolSAR image on FLEVOLAND area; (c) Classification map by the NRS method; (d) Classification map by the NRS\_MRF method; (e) Classification map by the weighted\_NRS method;  (f) Classification map by the proposed RNRS method without Tikhonov regularization term(noted by ``RNRS\_T"); (g) Classification map by the proposed RNRS method; (h) the corresponding land cover types for different colors.}
\label{fig1}
\end{figure*}

Our first data set is AIRSAR L band fully PolSAR image in Flevoland with the size of $300\times 270$. The Pauli RGB image and its groundtruth map are shown in Figs.\ref{fig1}(a) and (b).  In this image, there are 6 types of crops labeled in different colors, which are \emph{peas}, \emph{potatoes}, \emph{wheat}, \emph{barley}, \emph{beet} and \emph{bare soil} respectively. The white areas in the label map are unknown areas.

The experimental results by three compared and our methods are illustrated in Figs. \ref{fig1}(c)-(e) and(g) respectively. It can be seen that the three compared methods still produce some noisy classes even with ``3F+T" features. Considering the contextual information, the NRS\_MRF method can obtain  more homogenous regions, while some edge pixels are confused in Fig.\ref{fig1}(d). The proposed method shows better classification result in both edges and region homogeneity by utilizing RNRS model in Fig.\ref{fig1}(f).

To evaluate the proposed method quantitatively, the classification indicators of compared and proposed methods are given in Table \ref{flevd1}. It can be seen that our method can outperform others in  OA, AA and kappa indicators.  In addition, NRS\_MRF method shows compared quantitative result with our method. One reason is that the homogeneous regions are easy to be classified in Fig.\ref{fig1}. Compared with NRS\_MRF method, our method can also achieve better performance in relatively difficult heterogeneous regions as illustrated in Fig.\ref{fig2}. Second reason is that \textcolor{red}{the NRS\_MRF method uses} both the original PolSAR matrix and multiple of features (``3F+T") as the input, while our method only use the T matrix as the input. Multiple of features can improve the classification accuracy. It means that our method can obtain better results even with less features. Besides, our method can suppress speckle noises and obtain more homogeneous regions than the NRS\_MRF method from visual observation. Finally, the confusion matrix of proposed RNRS method is given in Table \ref{flevd2}.


Furthermore, to verify the effectiveness of the proposed Tikhonov regularization term $\Gamma _X^k$, the ablation experiment is conducted by adding the experimental result of the proposed RNRS method without Tikhonov regularization term(noted by ``RNRS\_T").  The experimental results by RNRS\_T and RNRS are given in  Figs. \ref{fig1}(f) and (g) respectively. Also, the classification indicators are given in Table \ref{flevd1}. It can be seen that the proposed method improve the classification accuracy obviously. The visual appearance is also improved by adding the Tikhonov regularization term in Fig.\ref{fig1} (g).

\begin{table}[ht]
\footnotesize
\begin{center}
\caption
{ \label{flevd1}
 Classification accuracy of different methods on FLEVOLAND Data Set(\%)}
\begin{tabular}{|p{1.1cm}|p{1.1cm}|p{1.1cm}|p{1.1cm}|p{1.1cm}|p{1.1cm}|}
\hline
class&NRS&NRS\_MRF &W\_NRS&RNRS\_T&RNRS\\
\hline
bare soil&81.49&82.75&84.86&86.36&\textbf{94.59}\\
potato&96.86&97.82& 90.88&\textbf{99.20}&98.68\\
beet&76.96&79.99&62.59&74.05&\textbf{80.63}\\
pea&96.72&96.59&96.51&96.98&\textbf{98.72}\\
wheat&97.59&\textbf{97.98}&93.953& 97.55&96.79\\
barley&97.14&\textbf{98.81}&87.21&98.37&96.21\\
OA&93.09&93.90&89.23&93.41&\textbf{95.15}\\
AA&91.16&92.29&86.12&92.68&\textbf{93.80}\\
Kappa&91.26&92.27&86.54&92.66&\textbf{93.85}\\
\hline
\end{tabular}
\end{center}
\end{table}

\begin{table}[ht]
\footnotesize
\begin{center}
\caption
{ \label{flevd2}
 Confusion matrix of the proposed method on FLEVOLAND Data Set}
\begin{tabular}{|p{1.0cm}|p{0.8cm}|p{0.8cm}|p{0.8cm}|p{0.8cm}|p{0.8cm}|p{0.8cm}|}
\hline
class&bare soil&potato &beet&pea&wheat&barley\\
\hline
bare soil&4790&0&0&0&274&0\\
potato&0&5451&1&0&0&72\\
beet&0&275&3591&96&0&664\\
pea&116&47&69&15648&187&111\\
wheat&94&93&0&6&9369&0\\
barley&0&64&95&0&0&5546\\
\hline
\end{tabular}
\end{center}
\end{table}

\subsection{Experimental Results of San Francisco Data Set}

\begin{figure*}
\centering
\setlength{\fboxrule}{0.2pt}
  \setlength{\fboxsep}{0.01mm}
\subfigure[]{\fbox{\includegraphics[height=0.14\textheight]{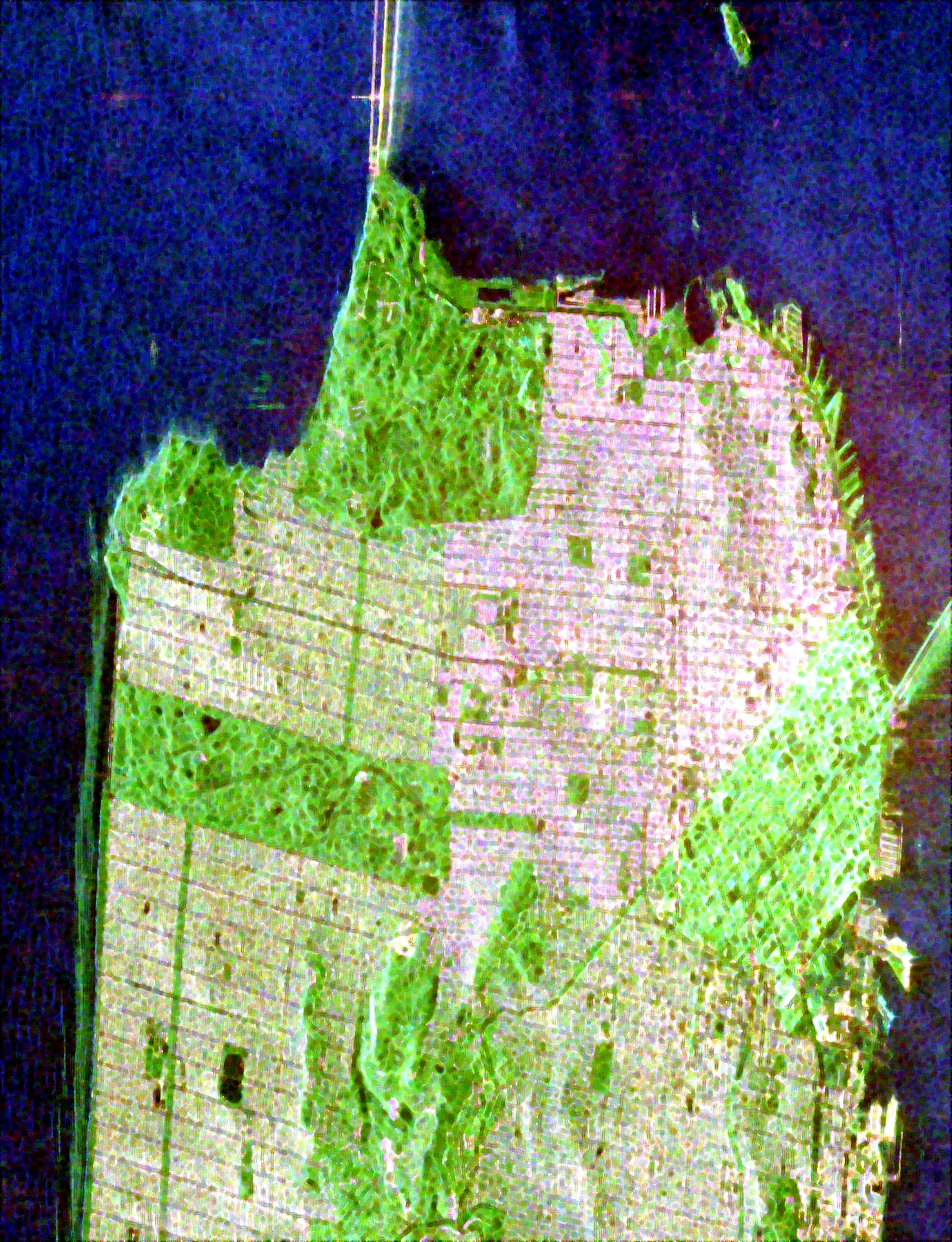}}}
\subfigure[]{\fbox{\includegraphics[height=0.14\textheight]{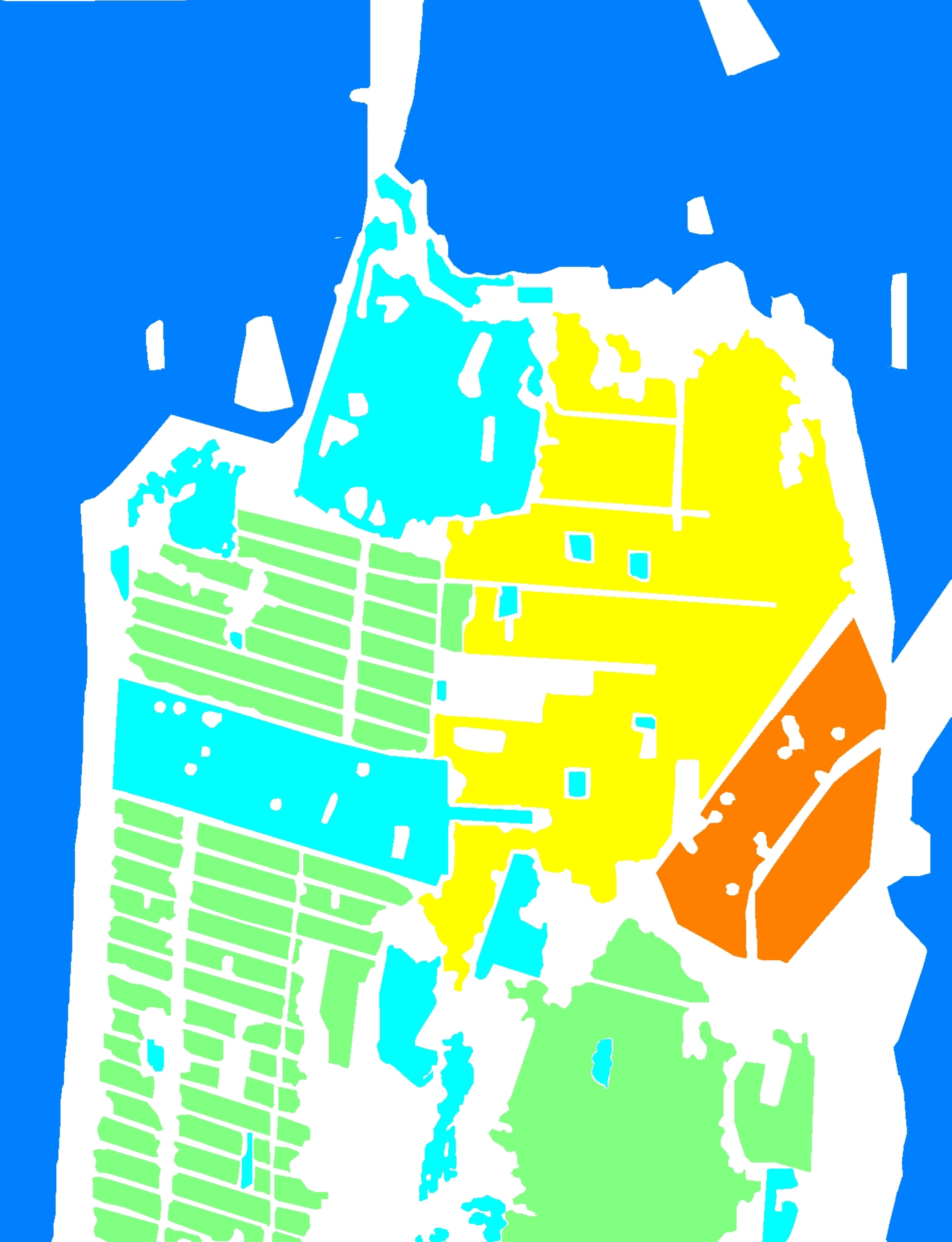}}}
\subfigure[]{\fbox{\includegraphics[height=0.14\textheight]{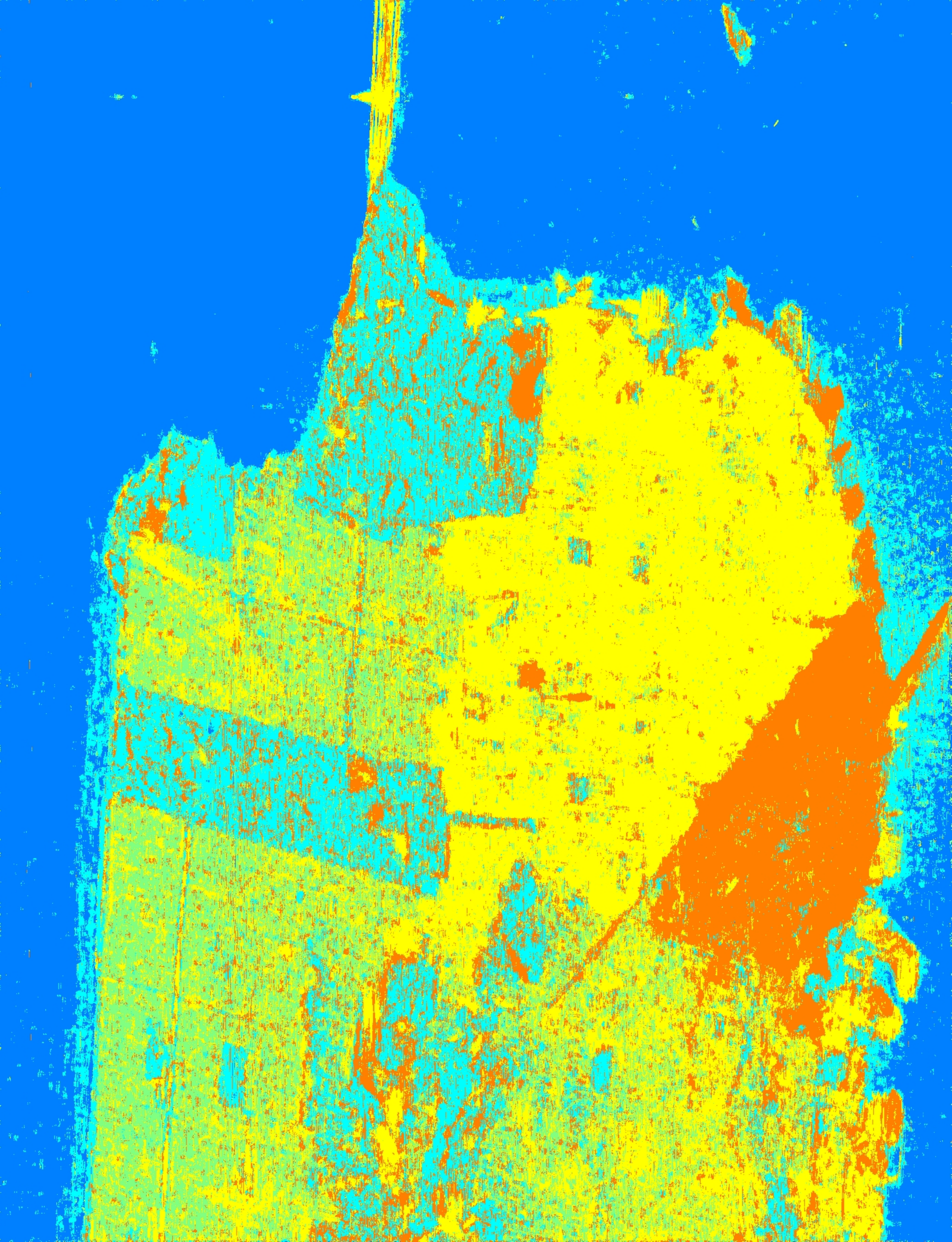}}}
\subfigure[]{\fbox{\includegraphics[height=0.14\textheight]{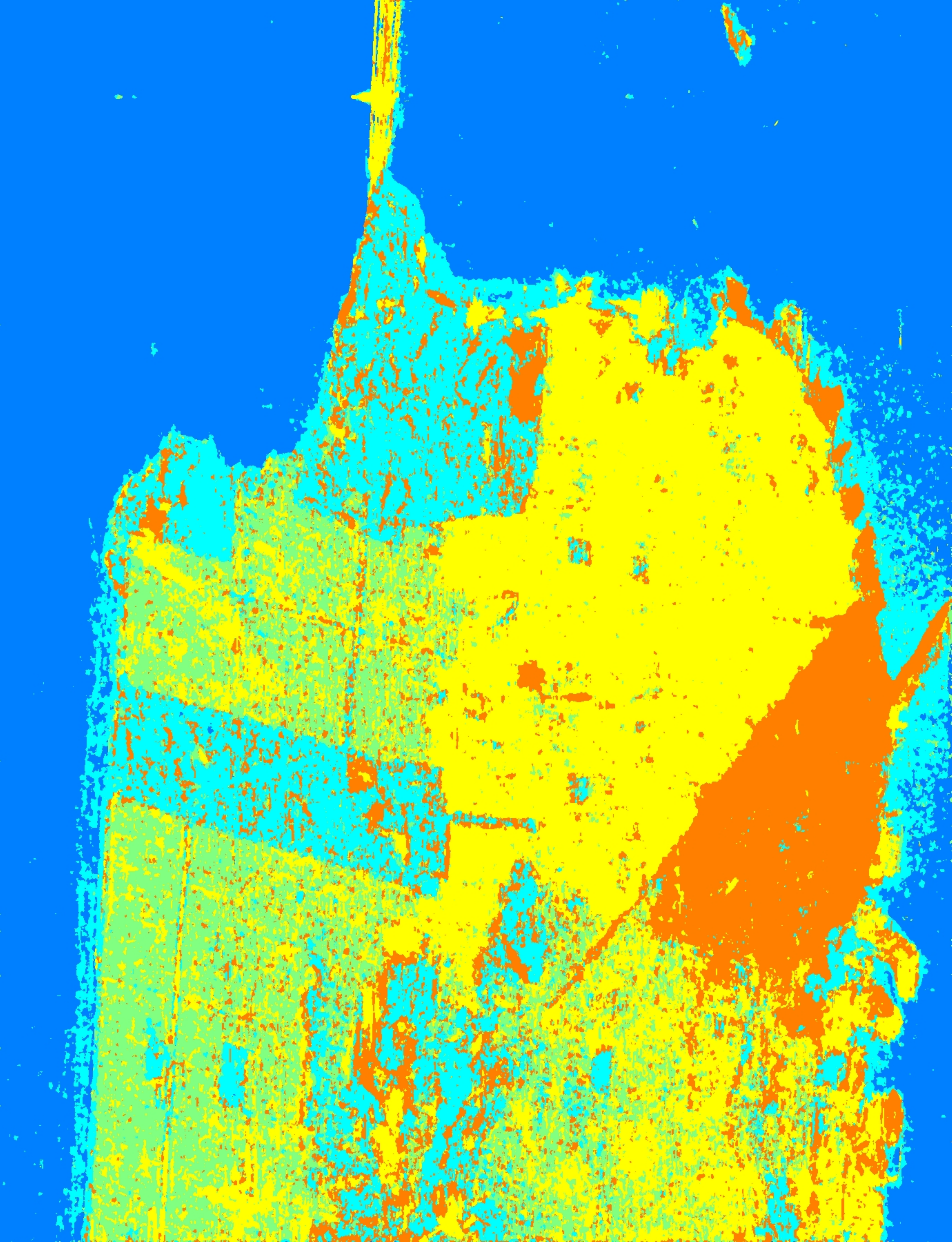}}}
\subfigure[]{\fbox{\includegraphics[height=0.14\textheight]{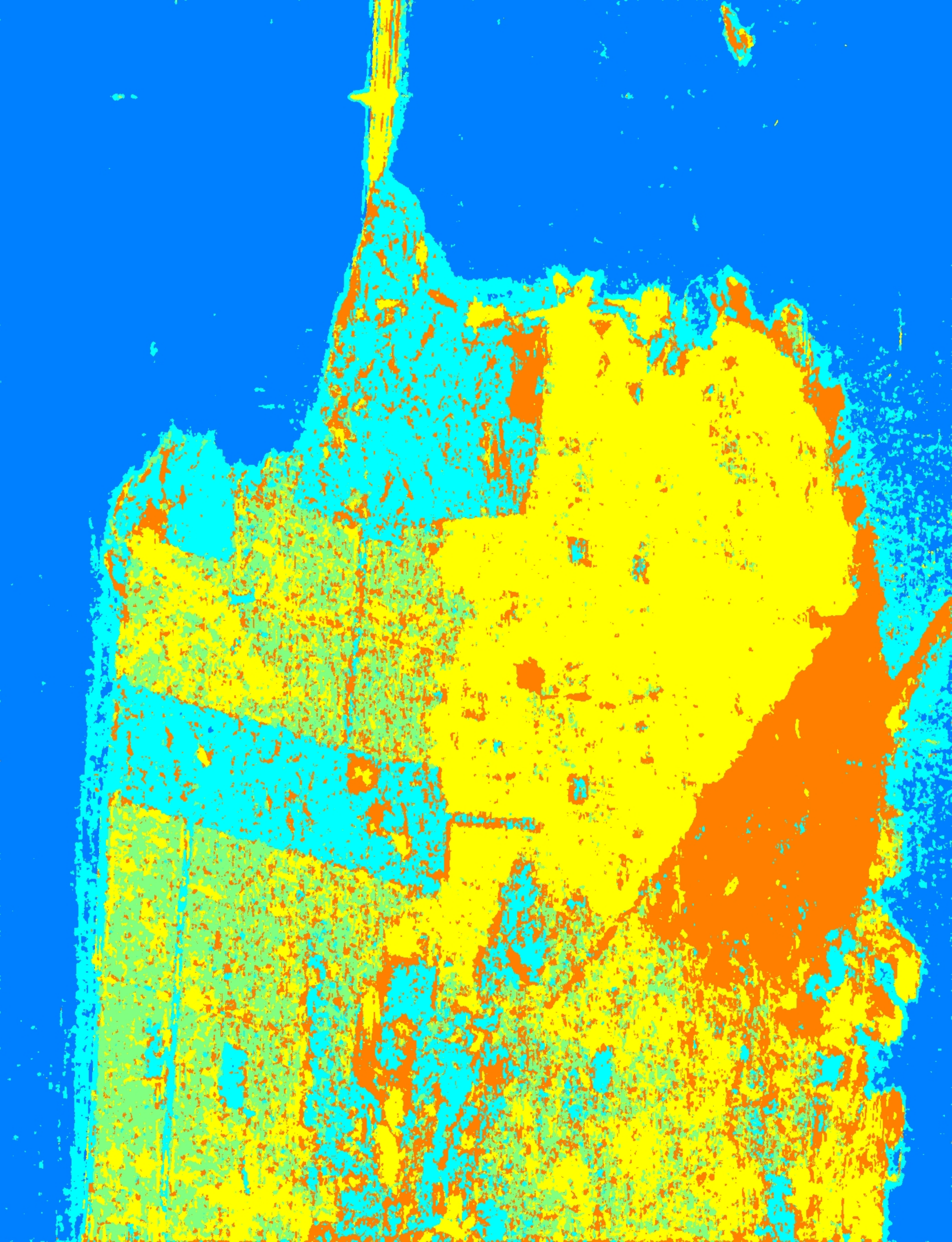}}}
\subfigure[]{\fbox{\includegraphics[height=0.14\textheight]{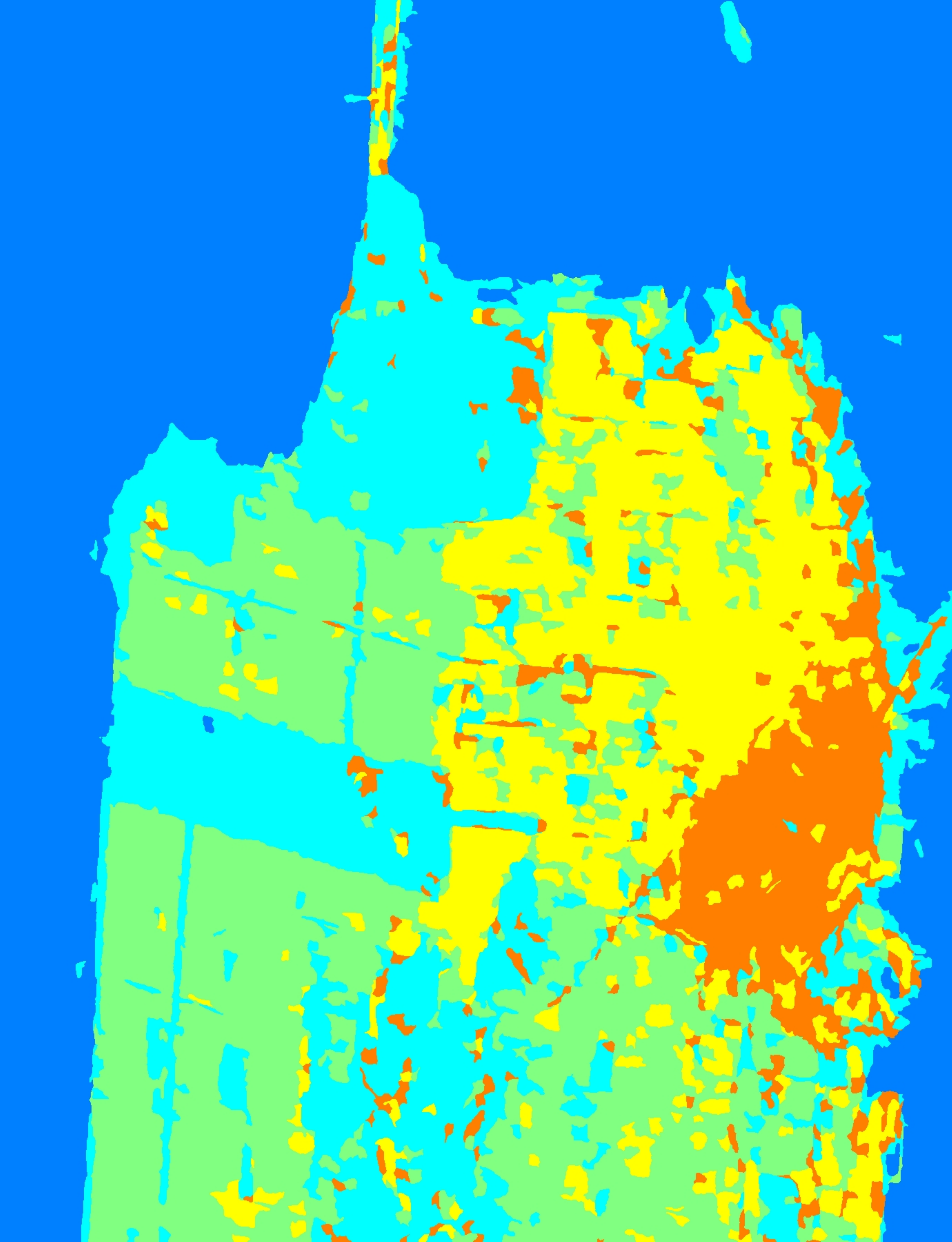}}}
\subfigure[]{\includegraphics[height=0.08\textheight]{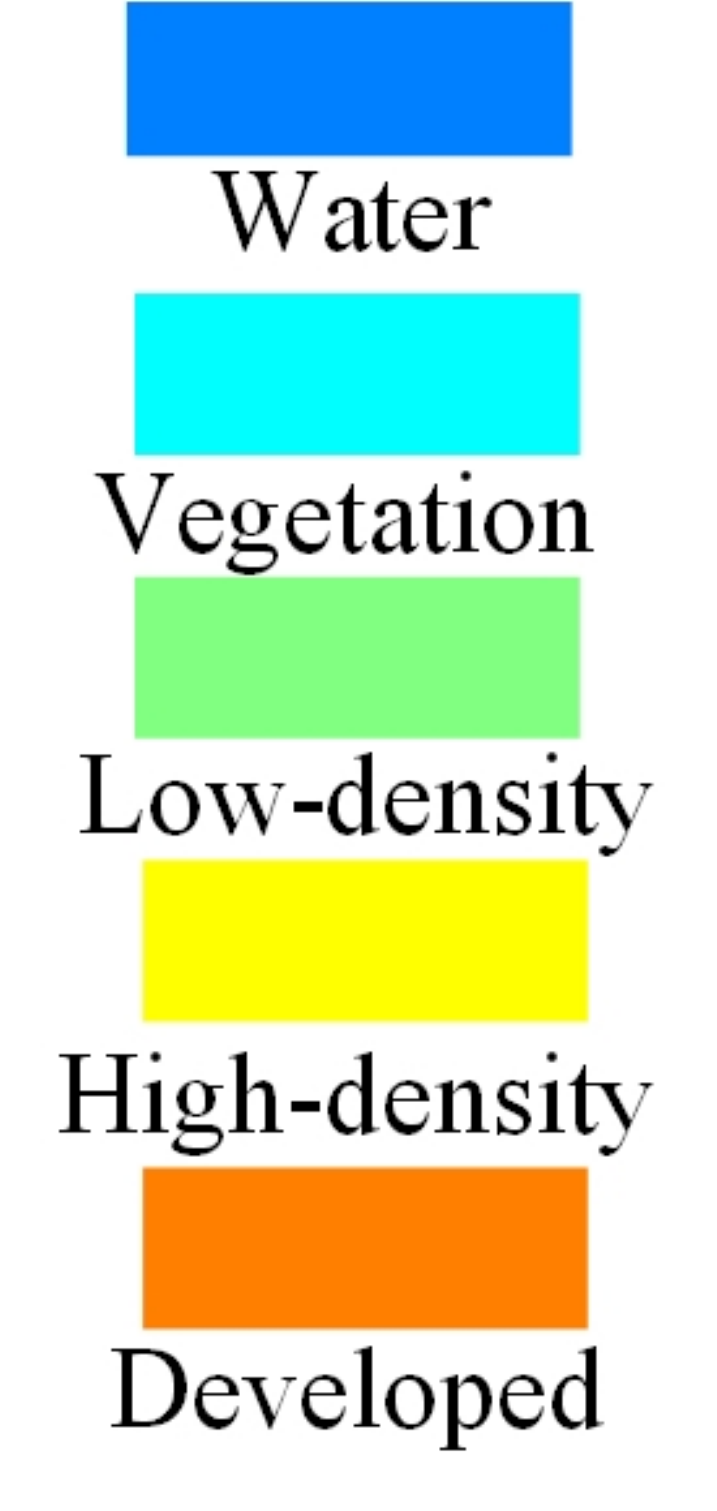}}

\caption{Classification results of San Francisco data set. (a) PauliRGB image of San Francisco area; (b) Ground truth of PolSAR image on San Francisco area; (c) Classification map by the NRS method; (d) Classification map by the NRS\_MRF method; (e) Classification map by the weighted\_NRS method; (f) Classification map by the proposed RNRS method; (g) the corresponding land cover types for different colors.}
\label{fig2}
\end{figure*}

Our second data set is RADARSAT-2 C band fully PolSAR image in San Francisco area. The image size is $1800\times1380$ pixels. The Pauli RGB image and its ground truth map are shown in Figs.\ref{fig2}(a) and (b) respectively. There are mainly five terrain types in this image, including the \emph{water}, \emph{vegetation}, \emph{low-density urban}, \emph{high-density urban} and \emph{developed}. Multiple of features(``3F+T'') are used in three compared methods, and ``only T" feature is applied to our method.

The experimental results by three compared and proposed methods are illustrated in Figs.\ref{fig2}(c)-(f). During the three compared methods, many noisy points are caused in the \emph{vegetation}, and they are misclassified into \emph{developed}. Besides, most pixels of the \emph{low-density} in the right bottom corner are misclassified into \emph{high-density}. Only the proposed method can classify the \emph{vegetation} and \emph{low-density} well. So, the proposed RNRS method can obtain more accurate classification map in (f). Moreover, the classification accuracy and confusion matrix are illustrated in Tables \ref{Sanfro1} and \ref{Sanfro2} respectively. Both of them show the advantages of the proposed method.

\begin{table}[ht]
\footnotesize
\begin{center}
\caption
{ \label{Sanfro1}
Classification accuracy of different methods on San Francisco Data Set(\%)}
\begin{tabular}{|p{1.5cm}|p{1.2cm}|p{1.2cm}|p{1.2cm}|p{1.2cm}|p{1.2cm}|}
\hline
class&NRS&NRS\_MRF &W\_NRS&RNRS\\
\hline
water&96.85&97.44&96.94&\textbf{99.63}\\
vegetation&61.20&69.07&73.20&\textbf{92.67}\\
low-density&45.22&49.28&40.77&\textbf{86.85}\\
high-density&92.25&\textbf{95.72}&95.21&78.55\\
developed&96.47&\textbf{98.56}&97.60&86.08\\
OA&81.37&84.12&	82.64&\textbf{92.31}\\
AA&78.40&82.21&80.74&\textbf{88.76}\\
Kappa&73.72&77.53&75.56&\textbf{88.96}\\	
\hline
\end{tabular}
\end{center}
\end{table}

\begin{table}[ht]
\footnotesize
\begin{center}
\caption
{ \label{Sanfro2}
 Confusion matrix of the proposed method on San Francisco Data Set}
\begin{tabular}{|p{1.2cm}|p{0.9cm}|p{1.0cm}|p{1.0cm}|p{1.0cm}|p{1.0cm}|}
\hline
&water&vegetation&low-density&high-density&developed\\
\hline
water&848893&3029&86&0&70\\
vegetation&591&219839&6000&2075&8732\\
low-density& 0&17926&305014&26930&1311\\
high-density&0&6684&44720&222288&9283\\
developed&0&53&0&11165&69398\\
\hline
\end{tabular}
\end{center}
\end{table}

\section{Conclusion}
This letter gave a  novel  Riemannian Nearest-Regularized Subspace classification method (RNRS), which used the original PolSAR matrix data as the input of the NRS method for the first time. Firstly, a Riemannian distance was designed to compute  the data loss term between real and estimated values. It made sure the model's input is the PolSAR original  matrix instead of feature vector.  Secondly, a  novel Tikhonov regularization term was designed to reduce the  pixels' distance within the same class. By these schemes, the proposed RNRS method learned the structure of original matrices  and correlation among channels effectively. Besides, the first-order derivation is inferred to solve the RNRS model. Experimental results showed the effectiveness of proposed method by comparing with other similar methods. It should be noted that all the compared methods use the multiple of features, including the scattering information, polarimetric data and image features. While our method use only the original PolSAR data feature. It verified the proposed method can obtain superior performance even with less features.


%

%
%

\ifCLASSOPTIONcaptionsoff
  \newpage
\fi



%

\bibliography{mybibfile}

\begin{thebibliography}{10}
\providecommand{\url}[1]{#1}
\csname url@samestyle\endcsname
\providecommand{\newblock}{\relax}
\providecommand{\bibinfo}[2]{#2}
\providecommand{\BIBentrySTDinterwordspacing}{\spaceskip=0pt\relax}
\providecommand{\BIBentryALTinterwordstretchfactor}{4}
\providecommand{\BIBentryALTinterwordspacing}{\spaceskip=\fontdimen2\font plus
\BIBentryALTinterwordstretchfactor\fontdimen3\font minus
  \fontdimen4\font\relax}
\providecommand{\BIBforeignlanguage}[2]{{%
\expandafter\ifx\csname l@#1\endcsname\relax
\typeout{** WARNING: IEEEtran.bst: No hyphenation pattern has been}%
\typeout{** loaded for the language `#1'. Using the pattern for}%
\typeout{** the default language instead.}%
\else
\language=\csname l@#1\endcsname
\fi
#2}}
\providecommand{\BIBdecl}{\relax}
\BIBdecl

\bibitem{2019A}
D.~Ratha, E.~Pottier, A.~Bhattacharya, and A.~C. Frery, ``A polsar scattering
  power factorization framework and novel roll-invariant parameter-based
  unsupervised classification scheme using a geodesic distance,'' \emph{IEEE
  Transactions on Geoscience and Remote Sensing}, vol.~58, no.~5, pp.
  3509--3525, 2020.

\bibitem{2020Object}
B.~Zou, X.~Xu, and L.~Zhang, ``Object-based classification of polsar images
  based on spatial and semantic features,'' \emph{IEEE Journal of Selected
  Topics in Applied Earth Observations and Remote Sensing}, vol.~PP, no.~99,
  pp. 1--1, 2020.

\bibitem{8994163}
J.~{Shi}, H.~{Jin}, and X.~{Li}, ``A novel multi-feature joint learning method
  for fast polarimetric sar terrain classification,'' \emph{IEEE Access},
  vol.~8, pp. 30\,491--30\,503, 2020.

\bibitem{8486711}
T.~{Dundar} and T.~{Ince}, ``Sparse representation-based hyperspectral image
  classification using multiscale superpixels and guided filter,'' \emph{IEEE
  Geoscience and Remote Sensing Letters}, vol.~16, no.~2, pp. 246--250, 2019.

\bibitem{8716570}
W.~{Li}, Y.~{Zhang}, N.~{Liu}, Q.~{Du}, and R.~{Tao}, ``Structure-aware
  collaborative representation for hyperspectral image classification,''
  \emph{IEEE Transactions on Geoscience and Remote Sensing}, vol.~57, no.~9,
  pp. 7246--7261, 2019.

\bibitem{6472065}
W.~{Li}, E.~W. {Tramel}, S.~{Prasad}, and J.~E. {Fowler}, ``Nearest regularized
  subspace for hyperspectral classification,'' \emph{IEEE Transactions on
  Geoscience and Remote Sensing}, vol.~52, no.~1, pp. 477--489, 2014.

\bibitem{Zhang2017Nearest}
Zhang, Fan, Jun, Yin, Qiang, Wei, Zheng, Liu, Yifan, and H.~and,
  ``Nearest-regularized subspace classification for polsar imagery using
  polarimetric feature vector and spatial information,'' \emph{Remote Sensing},
  2017.

\bibitem{2019Robust}
J.~Ni, F.~Zhang, Q.~Yin, and H.~C. Li, ``Robust weighting nearest regularized
  subspace classifier for polsar imagery,'' \emph{IEEE Signal Processing
  Letters}, vol.~PP, no.~99, pp. 1--1, 2019.

\bibitem{6918378}
L.~{Zhang}, L.~{Sun}, B.~{Zou}, and W.~M. {Moon}, ``Fully polarimetric sar
  image classification via sparse representation and polarimetric features,''
  \emph{IEEE Journal of Selected Topics in Applied Earth Observations and
  Remote Sensing}, vol.~8, no.~8, pp. 3923--3932, 2015.

\bibitem{7947120}
N.~{Zhong}, W.~{Yang}, A.~{Cherian}, X.~{Yang}, G.~{Xia}, and M.~{Liao},
  ``Unsupervised classification of polarimetric sar images via riemannian
  sparse coding,'' \emph{IEEE Transactions on Geoscience and Remote Sensing},
  vol.~55, no.~9, pp. 5381--5390, 2017.

\bibitem{1416405}
J.~A. {Tropp}, A.~C. {Gilbert}, and M.~J. {Strauss}, ``Simultaneous sparse
  approximation via greedy pursuit,'' in \emph{Proceedings. (ICASSP '05). IEEE
  International Conference on Acoustics, Speech, and Signal Processing, 2005.},
  vol.~5, 2005, pp. v/721--v/724 Vol. 5.

\bibitem{7729793}
J.~{Geng}, J.~{Fan}, H.~{Wang}, A.~{Fu}, and Y.~{Hu}, ``Joint collaborative
  representation for polarimetric sar image classification,'' in \emph{2016
  IEEE International Geoscience and Remote Sensing Symposium (IGARSS)}, 2016,
  pp. 3066--3069.

\bibitem{Anoop2017Riemannian}
Anoop, Cherian, Suvrit, and Sra, ``Riemannian dictionary learning and sparse
  coding for positive definite matrices,'' \emph{IEEE Transactions on Neural
  Networks and Learning Systems}, 2017.

\bibitem{IJCV}
F.~P. A.~N. Pennec, X., ``A riemannian framework for tensor computing,''
  \emph{Int J Comput Vision}, vol.~66, pp. 41--66, 2006.

\bibitem{6856946}
H.~{Song}, W.~{Yang}, X.~{Xu}, and M.~{Liao}, ``Unsupervised polsar imagery
  classification based on jensen-bregman logdet divergence,'' in \emph{EUSAR
  2014; 10th European Conference on Synthetic Aperture Radar}, 2014, pp. 1--4.

\bibitem{2001SPG}
``Spg: Software for convex-constrained optimization,'' \emph{ACM Transactions
  on Mathematical Software}, vol.~27, no.~3, pp. 340--349, 2001.

\bibitem{2020A}
J.~Shi, H.~Jin, and X.~Li, ``A novel multi-feature joint learning method for
  fast polarimetric sar terrain classification,'' \emph{IEEE Access}, vol.~PP,
  no.~99, pp. 1--1, 2020.

\end{thebibliography}

%
%

%

%
%
%




\end{document}